\documentclass[prl,twocolumn,preprintnumbers,amsmath,amssymb,superscriptaddress,nolongbibliography]{revtex4-1}
\usepackage{graphicx}% Include figure files
\usepackage{dcolumn}% Align table columns on decimal point
\usepackage{bm}% bold math
\usepackage{soul}

\usepackage{color}
\usepackage[normalem]{ulem}
\usepackage{hyperref,multirow}

\usepackage{float}
\usepackage{booktabs}
\usepackage{multirow}

\usepackage{ulem}
\usepackage[dvipsnames]{xcolor}

\usepackage{gensymb}

\usepackage{comment}

\begin{document}

\title{Link to Densify: Topological Transitions and Origin of Hysteresis During the (De)Compression of Amorphous Ices}

% Use letters for affiliations, numbers to show equal authorship (if applicable) and to indicate the corresponding author
\author{Yair Augusto Guti\'{e}rrez Fosado}
\affiliation{School of Physics and Astronomy, University of Edinburgh, Peter Guthrie Tait Road, Edinburgh, EH9 3FD, UK}
\thanks{yair.fosado@ed.ac.uk}
\author{Davide Michieletto}
\affiliation{School of Physics and Astronomy, University of Edinburgh, Peter Guthrie Tait Road, Edinburgh, EH9 3FD, UK}
\affiliation{MRC Human Genetics Unit, Institute of Genetics and Cancer, University of Edinburgh, Edinburgh EH4 2XU, UK}
\author{Fausto Martelli}
\affiliation{BM Research Europe, Hartree Centre, Daresbury WA4 4AD, UK}
\affiliation{Department of Chemical Engineering, University of Manchester, Manchester M13 9PL, UK}
\thanks{fausto.martelli@ibm.com}

% At least three keywords are required at submission. Please provide three to five keywords, separated by the pipe symbol.

\begin{abstract}
In this Letter we study the phase transition between amorphous ices and the nature of the hysteresis cycle separating them. We discover that a topological transition takes place as the system transforms from low-density amorphous ice (LDA) at low pressures, to high-density amorphous ice (HDA) at high pressures.  Specifically, we uncover that the hydrogen bond network (HBN) displays qualitatively different topologies in the LDA and HDA phases: the former characterised by disentangled loop motifs, while the latter displaying topologically complex and metastable Hopf-linked and knotted configurations. At the phase transition, the transient opening of the HBN topological motifs yields mechanical fragility on the macroscale. Our results provide a detailed microscopic description of the topological nature of the phase transition and the hysteresis cycle between amorphous ices. We argue that the topological transition discovered in this work may not only improve our understanding of amorphous ices but represent a generic mechanism for the densification of network-forming materials.
\end{abstract}

\maketitle

\paragraph{Introduction --} Water's distinctive molecular geometry and the capacity to form flexible hydrogen bonds (HBs), results in an extraordinarily intricate phase diagram encompassing, at least, 20 crystalline states~\cite{salzmann2019advances}, two liquid forms~\cite{palmer_nature,debenedetti2020,kim_2017,Nilsson2014}, and several amorphous phases~\cite{amann2016colloquium,gallo2016water,cassone2024electrofreezing} including the low-density amorphous ice (LDA) and the high-density amorphous ice (HDA)~\cite{mishima_1985,mishima_melting,hemley_new,mishima_relationship,debenedetti_2003,angell_2004,giovanbattista_2006,mishima_1998,loerting_the_2015,amann2016colloquium}. 
The hydrogen bond network (HBN) of LDA is mostly dominated by hexagonal motifs (or loops). When pressure is applied isotropically to LDA, the HBN configuration does not change considerably until a critical pressure $P_c^{*}$ is reached~\cite{formanek2020probing,formanek2023molecular}.
The demarcation line separating LDA and HDA seems to be of the first-order-kind~\cite{mishima_1985,mishima_1994,giovanbattista_2006,wong2015pressure,martelli2017large,martelli2018searching,giovambattista2021liquid,mollica2022decompression,formanek2023molecular,singh2024}, one of the most notable signatures being the characteristic hysteresis of the density during the compression and decompression. Several computational studies have looked at the kinetics of the transformation between these amorphous structures, providing important insights into the transition between them~\cite{wong2015pressure,martelli2017large,martelli2018searching,giovambattista2021liquid,mollica2022decompression,formanek2023molecular,singh2024}. However, a full understanding of the mechanism(s) underlying this phase transition is still lacking and has the potential to explain the microscopic origin of the hysteresis cycle.

Motivated by recent work on the liquid counterparts of LDA and HDA ices~\cite{Sciortino2022} and on the mechanical properties of DNA hydrogels~\cite{palombo2023topological} which display the emergence of topologically complex motifs including Hopf links and trefoil knots, here we explicitly look for complex topologies within the HBN of amorphous ices during compression and decompression. Our results provide a microscopic understanding of the transition between amorphous ices, explain the origin of the hysteresis separating them, and offer a topological explanation for the singular behavior of the isothermal compressibility. More broadly, our results provide a picture that may be the general mechanism of densification in network-forming amorphous materials including, e.g., silicon~\cite{deb2001pressure,mcmillan2005density,fan2024microscopic}, oxide glasses~\cite{meade1992high,kono2016ultrahigh}, chalcogenide glasses\cite{sen2006observation}, metallic glasses~\cite{sheng2007polyamorphism}, phase-change materials~\cite{sun2011pressure}, and organic materials~\cite{ha1996supercooled,wiedersich1997polyamorphism,zhu2017polyamorphism}, all of which are relevant to our daily life and broadly applied in industry. 

\begin{figure*}[t!]
	\centering
\includegraphics[width=0.95\textwidth]{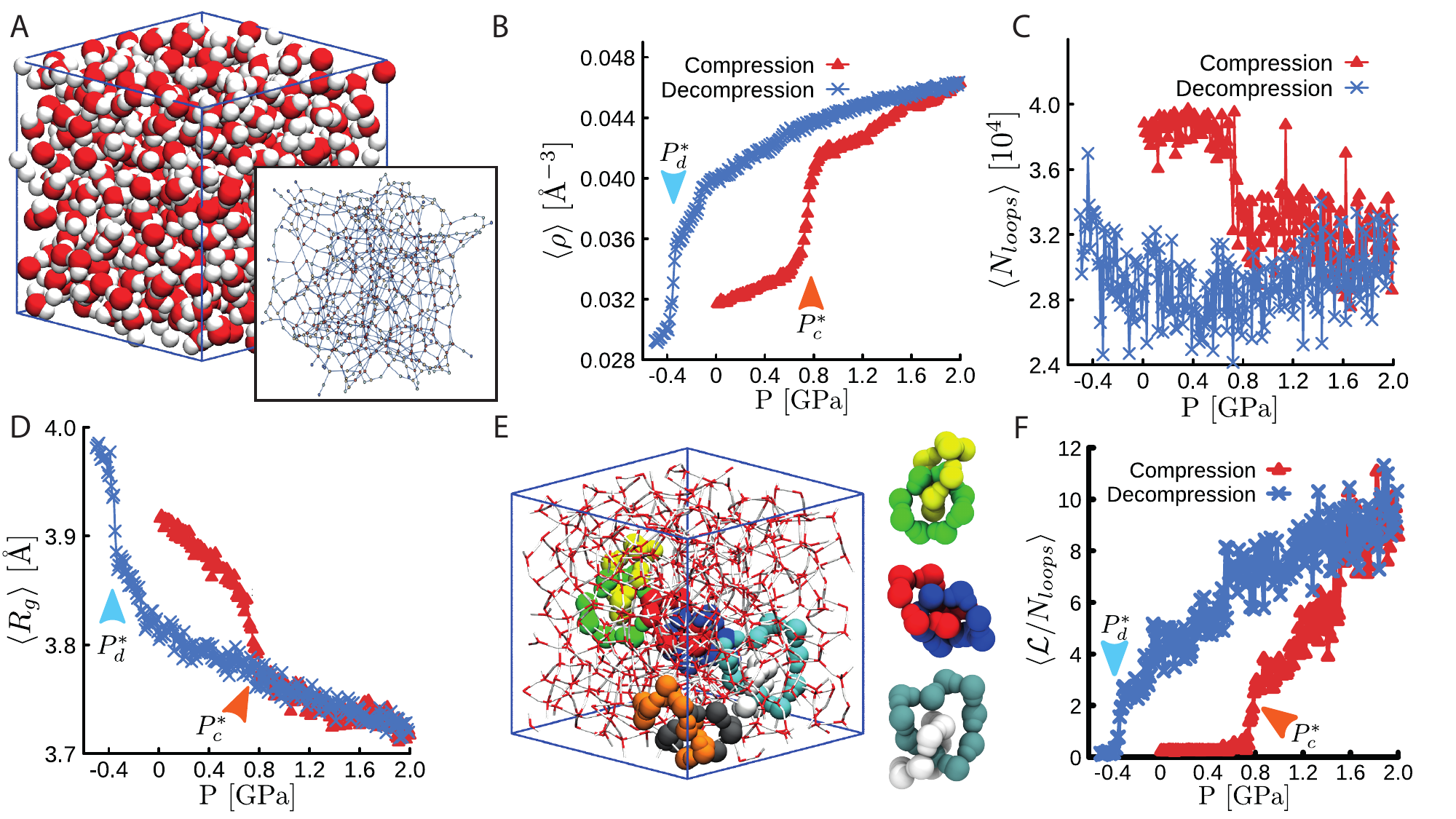}
	\vspace{-0.4 cm}
    \caption{ \textbf{Topological transition between LDA and HDA ice phases.} (A) Snapshot of the simulation box and (inset) corresponding graph representation of the HBN. (B) Average number density as a function of applied pressure. (C) The number of loops within the HBN (with a length shorter than 13 water molecules). (D) Average radius of gyration of the loops. (E) Snapshot of a simulation of HDA at 1~GPa where three pairs of Hopf-linked loops are highlighted. (F) Average valence, i.e. number of loops linked to any one loop within the HBN. Data points during compression (decompression) are reported as red (blue). All the data is obtained at $T=100$ K.}
	\label{fig:system}
 	\vspace{-0.4 cm}
\end{figure*}

\paragraph{Results --} To study the phase transition of amorphous ice, we performed classical molecular dynamics (MD) simulations of isothermal compression/decompression of water molecules modelled with the TIP4P/2005 interaction potential~\cite{tip4p2005}, which provides a reliable description of the phase diagram of amorphous ices~\cite{loerting2006amorphous}. We first generated LDA ice and simulated the compression/decompression cycles using the same protocols as in Refs.~\cite{wong2015pressure,martelli2018searching} and at temperatures $T=100$, $120$, and $140$ K (see SI). If not otherwise stated, we present results for samples with 512 water molecules and $T=100$ K (see SI for $T=120$, and $140$ K). The HBN is determined by looking for molecules whose oxygen atoms are closer than 2.2~\AA{} and for which the angle formed between O-H-O is smaller than 30$^{\circ}$~\cite{luzar1996hydrogen}. We then produced a graph where each vertex is a water molecule and each edge represents a hydrogen bond (HB, Fig.~\ref{fig:system}A). 

At $T=100$ K, amorphous ices described by the TIP4P/2005 water model undergo phase transitions during compression/decompression at $P_c^*\simeq 0.8$ and $P_d^* \simeq -0.4$ GPa, respectively~\cite{wong2015pressure,formanek2023molecular}. At these thermodynamic points, the average density displays an abrupt transition (Fig.~\ref{fig:system}B). To characterize the configurations acquired by the HBN we look for loops in the HBN graph representation as closed paths with no repetition of vertices or edges other than the starting and endpoints. 
We compute all the loops ($N_{i,l}$) formed by $l\in \left[ 3, l_{m} \right]$ water molecules, passing through the $i$th-vertex in the graph, $i\in \left[ 1, M=512 \right]$. We search for loops made at most by $l_{m}=13$ molecules (for computational feasibility, as done in Ref.~\cite{Sciortino2022}) and find the total number of loops as $N_{loops}=\sum_{i=1}^{M} N_{i,l}$. In Fig.~\ref{fig:system}C we report the average number of loops $\langle N_{loops}\rangle$ as a function of the applied pressure during compression/decompression. In LDA, $\langle N_{loops}\rangle$ fluctuates around $\langle N_{loops}\rangle \simeq 3.8\times10^{4}$, which is considerably larger than the amount of loops found recently in the LDA liquid counterpart~\cite{Sciortino2022}\footnote{Even more so because our system contains about half the number of molecules in Ref.~\cite{Sciortino2022}}. The fact that $\langle N_{loops}\rangle$ is roughly constant indicates that the HBN of LDA ice is mostly insensitive to this range of pressures~\cite{martelli2022steady,formanek2023molecular}. A sudden drop to $\langle N_{loops}\rangle \simeq 3.2\times10^{4}$ occurs at $P_{c}^* \simeq 0.8$ GPa, signaling the transition to HDA. Upon increasing further the pressure, the HBN is characterized by larger fluctuations of $\langle N_{loops}\rangle$, reflecting the sensitivity to the increased pressure characterizing HDA~\cite{formanek2023molecular}. The decompression of HDA shows an initial decrease in $\langle N_{loops}\rangle$ which can be attributed to the slower relaxation time of the HBN compared to the increased available volume. Eventually, $\langle N_{loops}\rangle$ increases reaching $\langle N_{loops}\rangle \simeq 3.2\times10^{4}$ at the lowest negative pressure, displaying a hysteresis cycle typical of first order phase transitions. 

\begin{figure*}[t!]
	\centering	\includegraphics[width=0.9\textwidth]{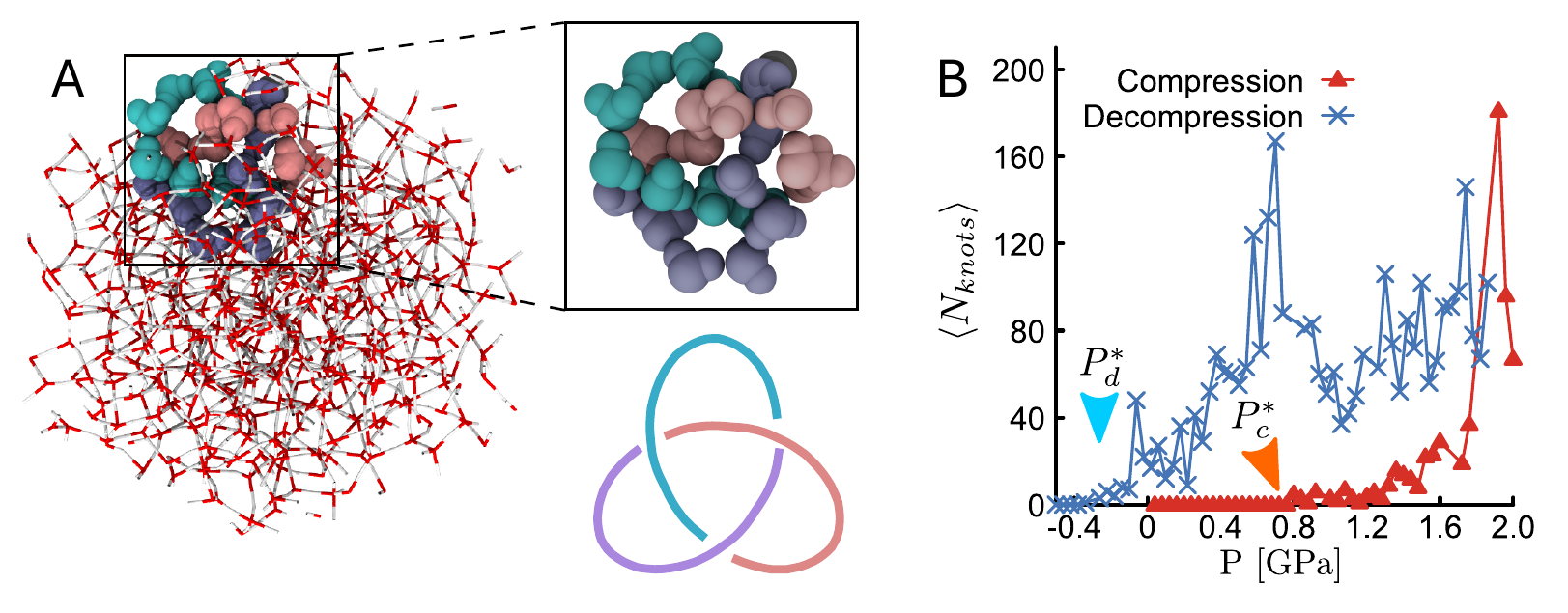}
 \vspace{-0.5 cm}
	\caption{ \textbf{Knots in HDA ice.} (A) Snapshot from simulations at $P = 2.0$~GPa and $T=100$~K. We highlight a trefoil knot found in our system among loops formed by 25 water molecules. (B) Average number of knots as a function of pressure during compression and decompression. Note that knots appear at the critical point $P_c^*$ and are metastable during decompression. }
	\label{fig:knotanalysis}
  \vspace{-0.3 cm}
\end{figure*}

In Fig.~\ref{fig:system}D we report the average radius of gyration $\langle R_{g} \rangle$ computed over all loops during compression/decompression. At $P < P_c^*$, the applied pressure rapidly decreases $\langle R_{g} \rangle$ and indicates that the loops in the HBN become increasingly smaller. Interestingly, they do not become shorter (see SI Fig.~S3), in turn implying that the HBN graph does not change topology significantly. 
At the critical pressure $P_c^{*}\simeq 0.8$ GPa, the behavior of $\langle R_{g} \rangle$ displays a sudden change in trend: higher pressures $P>P_c^*$ have a weaker effect on the size of the loops, as captured by the slower decay of $\langle R_{g} \rangle$. In seeming contradiction with this behaviour, the average length of the loops $\langle l \rangle$ steadily, albeit mildly, increases during compression (see SI, Fig.~S3). Upon decompression, the gyration radius of the loops remains metastable and returns to the original size only at the critical $P_d^* \simeq 0.4$ GPa, once again displaying a remarkable hysteresis cycle. 

Importantly, the seemingly contradictory observations that under compression the HBN forms \emph{smaller} yet mildly \emph{longer} loops, and that the system correspondingly displays an increases in density, can be understood as the emergence of interpenetrated HBN structures. Interpenetrated networks allow the system to reach higher densities by increasing the packing of water molecules yet by preserving the overall number of HBs. However, interpenetrated networks require longer loops to accommodate continuous paths and can form topologically complex and entangled structures such as knots and links~\cite{palombo2023topological}. 
To identify entangled motifs, we computed the linking number $Lk$ between pairs of closed-oriented loops $\gamma_{i}$ and $\gamma_{j}$ using the Gauss-Maxwell integral
\begin{equation}
Lk(\gamma_{i},\gamma_{j})=\frac{1}{4\pi} \displaystyle \oint_{\gamma_{i}} \oint_{\gamma_{j}} \frac{(\mathbf{r}_{j} - \mathbf{r}_{i})}{\lvert \mathbf{r}_{j} - \mathbf{r}_{i} \rvert^3} \cdot (\text{d}\mathbf{r}_{j} \times \text{d}\mathbf{r}_{i}),
\label{eq:Lk}
\end{equation}
where $\mathbf{r}_{i}$ and $\mathbf{r}_{j}$ are the vectors defining the position of all the points (the vertices on the graph) along the curves (the edges on the graph) $\gamma_{i}$ and $\gamma_{j}$, respectively. Examples of such topological links within the HBN are shown in Fig.~\ref{fig:system}E. 
%Since the orientation of the loops is randomly assigned, the value of $Lk$ averaged over all the pairs of loops is close to zero. 
To compute the overall degree of linking within the HBN, we compute the total number of times that two loops are linked regardless of the sign of $Lk$, i.e. $\mathcal{L} = \sum_{i>j}^{N_{loops}} \lvert Lk(\gamma_{i},\gamma_{j}) \rvert$. 

The quantity $\mathcal{L}$, scaled by the average number of loops in the HBN provides the so called ``valence'', or the number of links for any one loop in the HBN (see Fig.~\ref{fig:system}F). While the LDA ice phase is characterized by the absence of links between the loops, one can appreciate the abrupt transition at $P_c^{*}$, which marks the stability limit of the LDA. Above this point, the network's configurational entropy increases thanks to the interpenetration and interlinking of loops, via the formation of simple Hopf links.

Both experiments~\cite{shephard2016molecular} and simulations~\cite{formanek2023molecular} have reported the partial activation of molecular rotations in correspondence with $P_c^*$. The reorientation of water molecules modifies the HBN topology by allowing the transient opening of loops characterizing the LDA phase, and by creating new entangled motifs that can accommodate the increased density of HDA. Above $P_c^*$, the average valence $\langle\mathcal{L}/N_{loops}\rangle$ displays a rapid increase (Fig.~\ref{fig:system}F): indeed, the fast increase in linking between loops and the slow decrease in their size ($\langle R_{g} \rangle$, Fig.~\ref{fig:system}D) indicates that the densification of the HBN is mostly driven by the increase in topological complexity rather than the reduction in loop size. Thus, we argue that Hopf-linked motifs underlie the hypothesized connection between HDA and ice IV, itself characterized by interpenetrating loops~\cite{shephard2017high,martelli2018searching}. 
The decompression of HDA induces an increase of $\langle R_{g} \rangle$ (Fig.~\ref{fig:system}D) and is accompanied by a slow reduction of $\langle\mathcal{L}/N_{loops}\rangle$ (Fig.~\ref{fig:system}F). In turn, this suggests that the links between the loops are stable, yet the loops are less distorted. At the decompression critical pressure, $P_d^{*}\simeq -0.4$ GPa, $\langle R_{g} \rangle$ undergoes a sudden increase simultaneously with a sudden drop in $\langle\mathcal{L}/N_{loops}\rangle$. The critical temperature $P_d^*$ is characterized by the partial activation of molecular rotations~\cite{shephard2016molecular,formanek2023molecular} which allow linked motifs to disentangle to recover the LDA phase. 

\begin{figure*}[t!]
	\centering
	\includegraphics[width=0.95\textwidth]{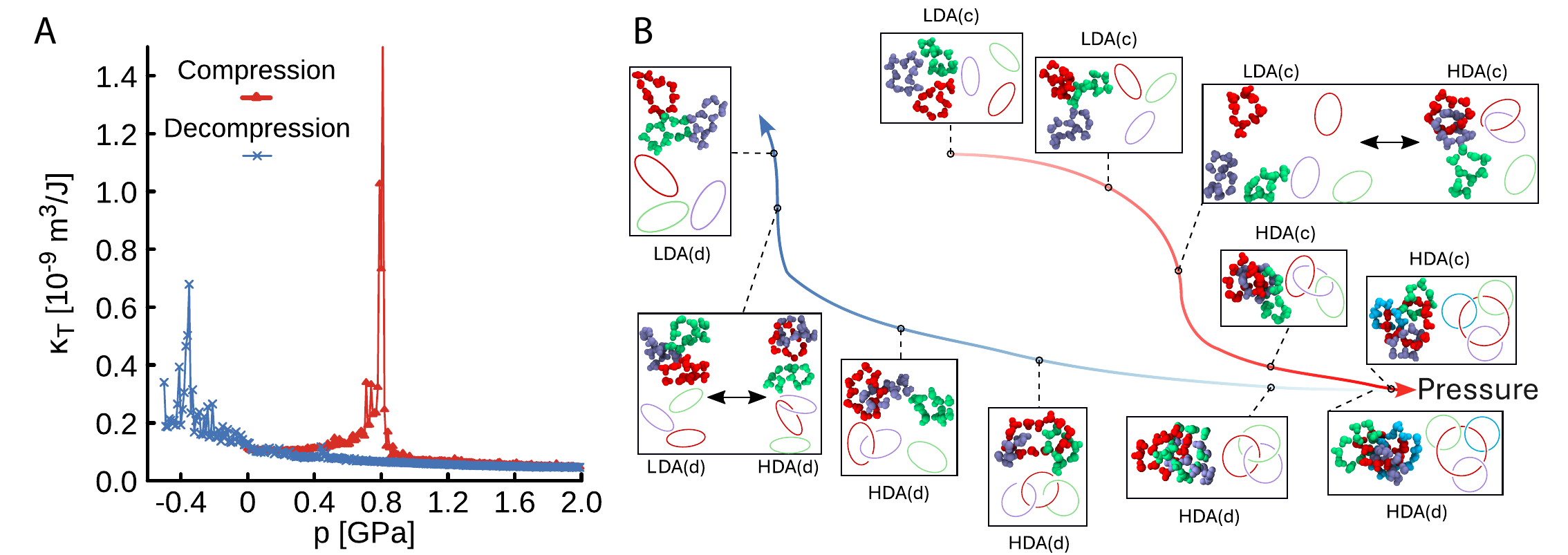}
    \vspace{-0.3 cm}
	\caption{\textbf{Effects of topology on compressibility of amorphous ices.} (A) Profile of the isothermal compressibility $k_T$ computed at $T=100$~K during compression (red symbols) and decompression (blue symbols). (B) Pictorial representation of the HBN rearrangement during compression and decompression of amorphous ices. Snapshots of loops within the HBN are shown along the curves. }
     \vspace{-0.3 cm}
	\label{fig:compressibility}
\end{figure*}

%Notably, in the set of loops with 13 or fewer water molecules, we did not find links with $|Lk|>1$. 
Motivated by the discovery of links, we asked if our systems displayed knot topologies. We thus searched for knotted states of individual loops using KymoKnot~\cite{kymoknot} and could not find knots formed by loops with 13 or fewer water molecules. 
%At our highest pressure (2 GPa), knots could only started to be found within loops with $21$ water molecules, for which we identified a total of $\sim 2\times10^7$ loops and only 4 knots (all trefoils). 
%Therefore, the average number of knots per loop is considerably smaller than the average number of links per loop. 
To generate a more complete statistics of knot formation, we then searched through all the loops made of exactly $25$ water molecules and found that the system displayed up to $\sim 200$ knots at our highest pressure (Fig.~\ref{fig:knotanalysis} A). This number is more than 100-fold smaller than the number of links in the system. The search of knots composed up to 25 water molecules is computationally prohibitive. Importantly, and mirroring the behaviour of Hopf links, during compression the appearance of knots occurs at the critical pressure $P_c^*$, and they remain metastable during decompression until the critical $P_d^*$ (Fig.~\ref{fig:knotanalysis}B).

We further asked oursevles if the abrupt change in HBN topology observed at $P_c^*$ and $P_d^*$ could have a measurable effect on the physical and mechanical properties of our system. In Fig.~\ref{fig:compressibility}A, we report the isothermal compressibility $\kappa_T$ (which is the inverse of the bulk modulus), computed during compression and decompression. Away from the phase transitions, the samples are characterized by low values of $\kappa_T$, indicating that the HBN acquires configurations capable of efficiently absorbing long-range density fluctuations~\cite{martelli2017large,formanek2023molecular}. On the other hand, the compressibility markedly increases in correspondence with the phase transitions. Such loss of mechanical rigidity results from the transient opening of loops, which creates a HBN unable to absorb long-range density fluctuations and sustain the externally applied pressure. The opening of the loops to allow for the creation and the resolution of Hopf-linked loops therefore explains the observed loss of hyperuniformity in these samples in correspondence with the phase transitions~\cite{martelli2017large,formanek2023molecular}. Nonetheless, despite its complexity, the HBN topology of HDA creates mostly incompressible configurations at the limit of nearly hyperuniformity~\cite{martelli2017large,formanek2023molecular}. Unveiling the intricacies of this network adds a dimension to the realization of disordered hyperuniform materials.

%Our results indicate that the configurational entropy of the HBN is the driving force in the transitions between LDA and HDA during the compression/decompression cycles. The initial increase in density in LDA occurs via the deformation of the motifs populating the HBN. In this case, the applied pressure is not high enough to allow for interlinking. The transition to HDA occurs only when the pressure is intense enough to overcome the stability of the HBN and increase the configurational entropy via the introduction of Hopf links. The appearance of Hopf links occurs through the momentary opening of disentangled loops which causes a loss of rigidity in the sample. The consecutive closing of the loops generates Hopfs links. These entangled motifs may be the source of the signature of the interpenetrating network of metastable ice IV previously observed in experiments~\cite{shephard2017high} and simulations~\cite{martelli2018searching}. The decompression of HDA induces an increasingly available space allowing for the full disentanglement of the HBN only at negative pressures, with the disappearance of Hopf links and the recovery of LDA. The disentanglement of the HBN occurs via the rupture of loops which, similarly to the transition from LDA to HDA, induces a loss of rigidity in the sample.\newline 

\paragraph{Conclusions. --}
In this Letter we have discovered that amorphous ices undergo a first-order phase transition that involves the formation of topologically complex motifs within the HBN. The emerging picture can be summarized as follows (see Fig.~\ref{fig:compressibility}B): for $P<P_c^*$ the isotropic pressure shrinks the loops of the HBN without affecting their connectivity. At $P\sim P_c^*$, molecular rotations allow loops within the HBN to transiently open and then re-form into Hopf-linked motifs which accommodate a higher density of water molecules. The number of topologically complex motifs such as Hopf links (Fig.~\ref{fig:system}), and knots (Fig.~\ref{fig:knotanalysis}), increases abruptly at $P_c^*$ in correspondence to the transition to HDA ice phase. During decompression, these topologically complex motifs remain metastable through the critical point, and the full disentanglement of the HBN motifs is recovered only at negative pressures $P_d^*$ in the LDA ice phase. 

We note that the HBN of LDA and HDA ice phases host topological motifs similar to those of their corresponding liquid phases~\cite{Sciortino2022}. This supports the conjecture that there exist a liquid-liquid critical point for this water model and that LDA and HDA phases may be the glassy equivalents of the equilibrium liquids~\cite{martelli2020connection}. 
We thus argue that the recently reported medium-density amorphous (MDA) ice~\cite{rosu2023medium,eltareb2024continuum} should be investigated in terms of the complex topological motifs inspected in this work and Ref.~\cite{Sciortino2022}. Our results and the structural similarities recently found between some of the MDAs and LDA phases~\cite{faure2024high} suggest that MDAs may correspond to LDA or HDA configurations with distorted or interpenetrating HBNs. 

Our study explains the topological nature and origin of the hysteresis cycle between amorphous ices and the first-order phase transition separating them. The network's topological re-arrangement uncovered in this work may be a general mechanism of densification in amorphous network-forming materials at large, including semiconductors and pharmaceuticals. 

\paragraph{Acknowledgements --}
F.M. acknowledges support from the Hartree National Centre for Digital Innovation, a collaboration between STFC and IBM. DM acknowledges the Royal Society and the European Research Council (grant agreement No 947918, TAP) for funding. The authors also acknowledge the contribution of the COST Action Eutopia, CA17139. For the purpose of open access, the author has applied a Creative Commons Attribution (CC BY) licence to any Author Accepted Manuscript version arising from this submission. 
Codes are available at: \href{https://git.ecdf.ed.ac.uk/ygutier2/amorphous-ice-topology}{\small{https://git.ecdf.ed.ac.uk/ygutier2/amorphous-ice-topology}}

\bibliographystyle{apsrev4-1}
\bibliography{bibliography}

\renewcommand{\thefigure}{S\arabic{figure}}
\setcounter{figure}{0}

\newpage

\section{Supplementary Information}

\section{Molecular dynamics simulations}
Classical molecular dynamics simulations were performed on $N=512$ rigid water molecules described by the TIP4P/2005 interaction potential~\cite{tip4p2005} in the isobaric ($N$$P$$T$) ensemble using the \textsc{GROMACS 2021.5} package~\cite{gromacs}. The TIP4P/2005 water model reproduces well the phase diagram of water at the thermodynamic conditions of interest of this work~\cite{wong2015pressure,martelli2018searching}. Coulombic and Lennard-Jones interactions were calculated with a cutoff distance of $1.1$~nm and long-range electrostatic interactions were treated using the Particle-Mesh Ewald (PME) algorithm. Temperatures and pressures are controlled using the Nos\'e-Hoover thermostat~\cite{nose,hoover} with a constant of $0.2$~ps, and the Berendsen barostat~\cite{berendsen1984molecular} with a time constant of $1$~ps. Equations of motions are integrated with the Verlet algorithm~\cite{verlet} with a time step of $1$~fs. \newline 
The simulation protocol follows previous works employing the same water model~\cite{wong2015pressure,martelli2017large,martelli2018searching}. LDA configurations were obtained by cooling liquid water equilibrated at T=300~K with a quenching rate of 1~K/ns down to T=100~K. Compression/decompression cycles were simulated with a compression/decompression rate of 0.01~GPa/ns at T=100~K, T=120~K and T=140~K. The highest pressure simulated was 2~GPa, the lowest was -0.5~GPa.

\section{Links and Knots}
\begin{figure}[t!]
	\centering
	\includegraphics[width=0.45\textwidth]{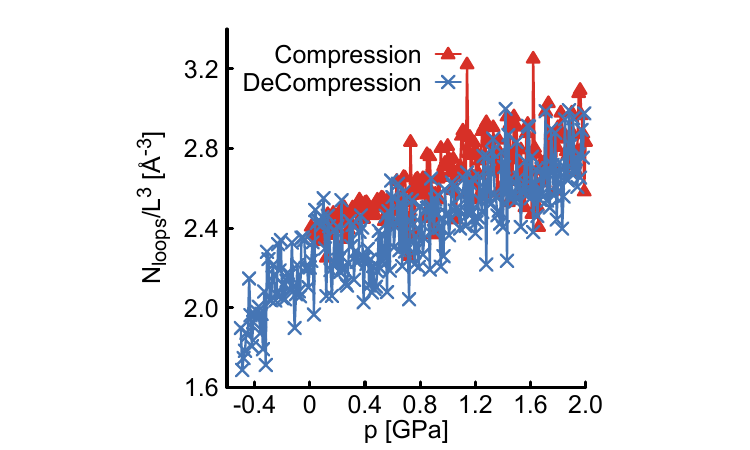}
	\caption{Density of loops ($N_{loops}/L^{3}$, with $L$ the box size) as function of the pressure at T=100 K. During compression $N_{loops}/L^{3} \in [2.2, 3.0] \AA^{-3}$, while during decompression $N_{loops}/L^{3} \in [1.6, 2.8] \AA^{-3}$}
     %\vspace{-0.6cm}
	\label{fig:loopdens}
\end{figure}

To improve the efficiency of our analysis, after identifying all the $N_{i,l}$ loops involving a vertex $i$, we remove that vertex and continue our search at vertex $i+1$. 
To prevent the identification of linear paths as loops (closed through the periodic boundary conditions), we use the minimum image criterion (MIC) to reconstruct the loops, and we impose the condition on each loop's radius of gyration $R_{g}<L/3$, with $R^{2}_{g}=\dfrac{1}{l}\sum_{n=1}^{l}[\mathbf{r}_{\text{mean}}-\mathbf{r}_{n}]^{2}$. In this relation, $\mathbf{r}_{n}$, $\mathbf{r}_{\text{mean}}$ and $l$ represent the position of the n-th oxygen forming the loop, the center of mass of the loop and its size, respectively. In Fig.~\ref{fig:loopdens} we report the density of loops as function of the pressure. We note that on average the density of loops is conserved (also at the transition pressures $P_{c}^{*}=0.8$ GPa and $P_{d}^{*}=-0.4$ GPa) 

The identification of knotted states of individual loops (performed by using KymoKnot~\cite{kymoknot}), requires the use of longer loops than the computation of the linking $\mathcal{L}$. Our analysis revealed that at our highest pressure (2 GPa), knots could only be found when $l_{max} \geq 21$. In fact, for $l_{max}=21$ we identified a total of $\sim 2\times10^7$ loops and only 4 knots (all of them trefoils). Therefore, the average number of knots per loop is considerably smaller than the average number of links per loop. Although it was possible to find knots at $P=2$ GPa when $l_{max}=21$, this was not enough to capture the discontinuity in the number of knots as a function of the pressure, that is expected to signal a first-order phase transition. Since the number of loops increases exponentially with $l_{max}$, the analysis becomes computationally expensive and unfeasible for $l_{max}>21$. 
One possible way to overcome this technical issue is by finding only rings made of exactly (and not up to) $l^{*}=25$ water molecules. In addition, we set 500000 as the maximum number of rings that our algorithm can find passing through a single vertex in the graph. With these two conditions, we computed the number of knots ($N_{knots}$) as a function of the pressure reported by the red data points in Fig.~\ref{fig:knotanalysis}B, when the system is under compression. In this plot, we can identify the critical pressure ($P_{c}=0.8$ GPa) at which the system undergoes a change in its topological state (from unknotted to knotted), but the analysis does not provide a full description about the knotting probability of the system. Analogous results during decompression are reported by the blue data points in Fig~\ref{fig:knotanalysis}(b, right), where we identify $P_{d}=-0.4$ GPa, in agreements with the transition observed with $\mathcal{L}$.

\begin{figure}[t!]
	\centering
	\includegraphics[width=0.4\textwidth]{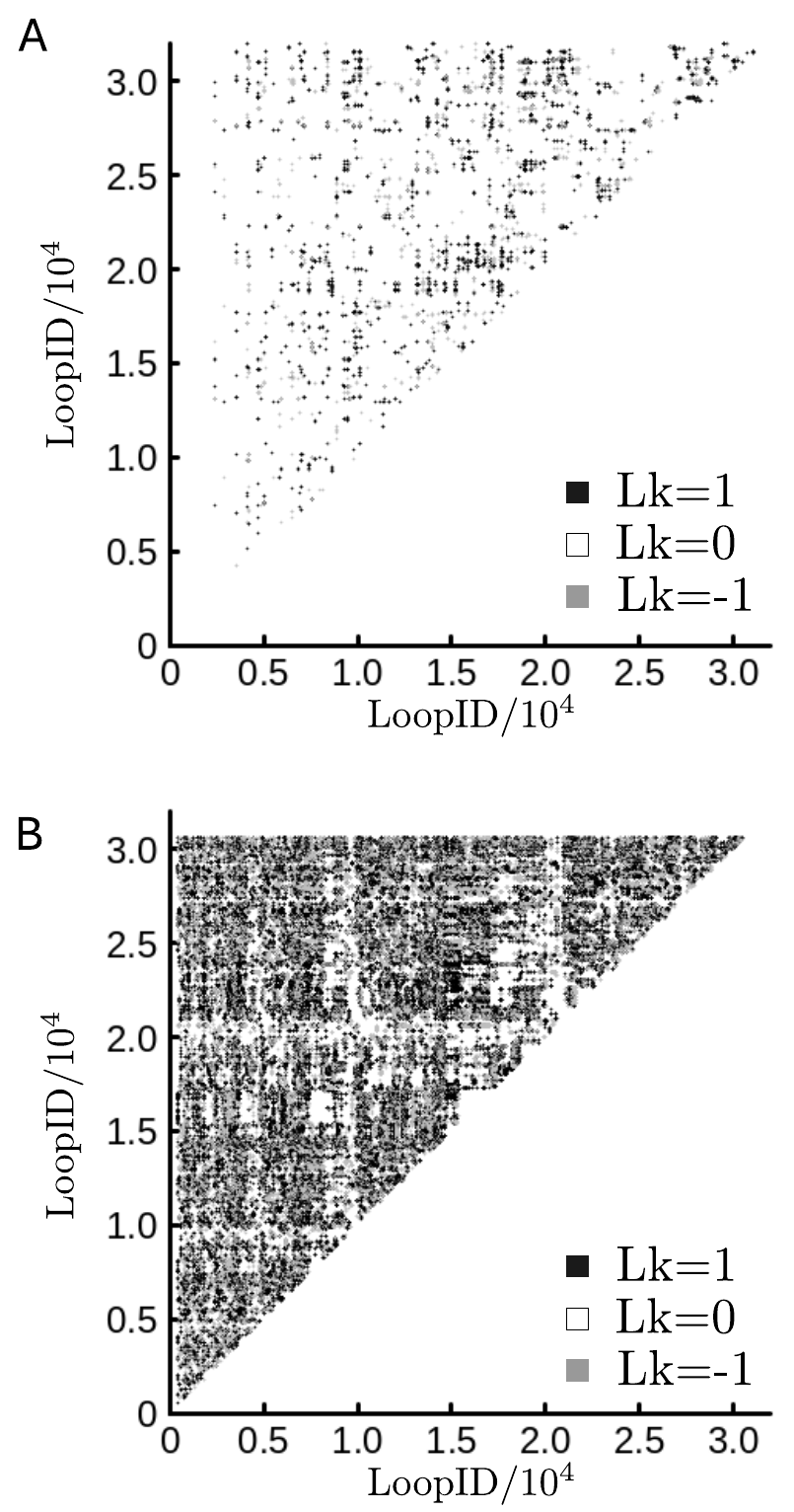}
	\caption{Linking matrix computed at P=0.4~GPa (A) and P=1.6~GPa (B).}
     %\vspace{-0.6cm}
	\label{fig:links}
\end{figure}

We obtained the linking matrix in Fig.~\ref{fig:links} by filling the $m,n$ element of the matrix with the value of the linking number between the pair of loops with ID=$m$, $n$. We report results at T=100 K and at two different pressures (during compression), corresponding to LDA (P=0.4~GPa, panel A) and HDA (P=1.6~GPa, panel B). We note that in the linking matrix we do not observe elements with $\lvert Lk \rvert > 1$.

\begin{figure}[t!]
	\centering
	\includegraphics[width=0.48\textwidth]{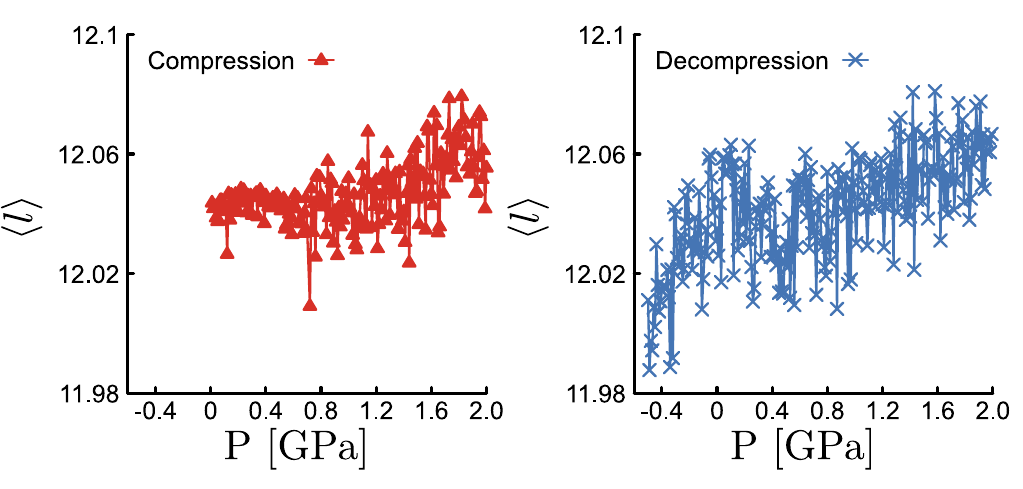}
	\caption{Average length of the topological motifs during compression and decompression.}
     %\vspace{-0.6cm}
	\label{fig:length}
\end{figure}

\begin{figure}[t!]
	\centering
	\includegraphics[width=0.45\textwidth]{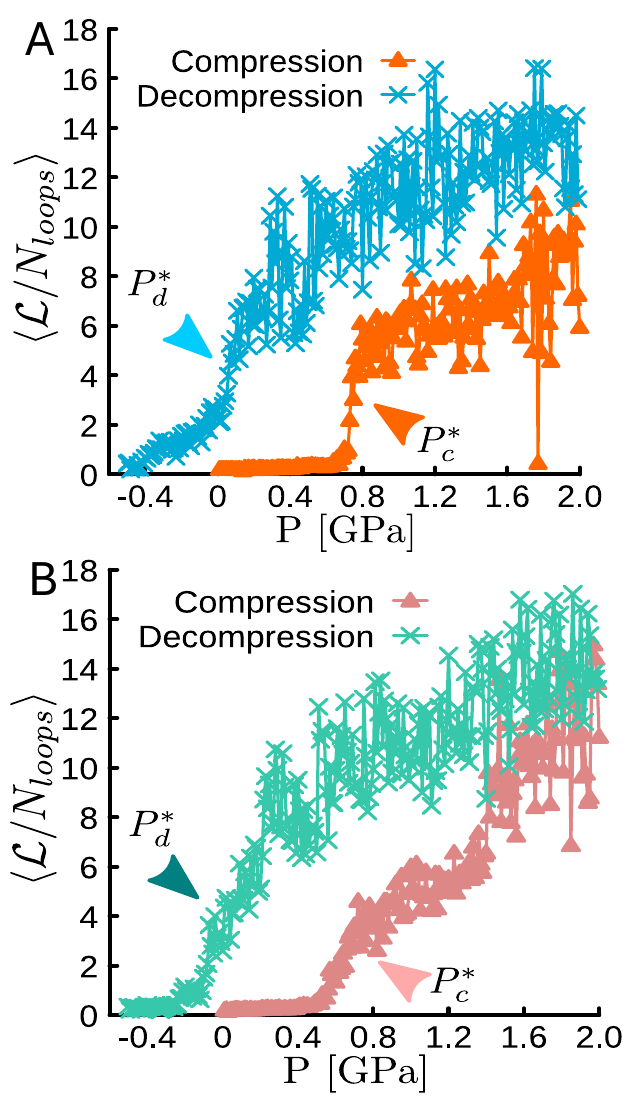}
	\caption{$\langle \mathcal{L}/N_{loops} \rangle$ computed during the compression/decompression cycles at T=120~K (A) and T=140~K (B).}
     %\vspace{-0.6cm}
	\label{fig:loops}
\end{figure}

In Fig.~\ref{fig:loops} we report $\langle \mathcal{L}/N_{loops} \rangle$ computed at higher temperatures compared to the main text. Panel A refers to T=120~K, while panel B refers to T=140~K.

\end{document}